\documentclass[prl,article,twocolumn,reprint,superscriptaddress]{revtex4-2}
\usepackage{graphicx,amsfonts,color,comment,amsmath,hyperref,float}
\usepackage{dcolumn}% Align table columns on decimal point
\usepackage{bm}% bold math
\usepackage{url}
\usepackage{soul}
\begin{document}
%\preprint{APS/123-QED}

\title{Evidence for PeV Proton Acceleration from \\ Fermi-LAT Observation of SNR G106.3+2.7 }% Force line breaks with \\
 
\author{Ke Fang}
 \affiliation{Department of Physics, Wisconsin IceCube Particle Astrophysics Center, University of Wisconsin, Madison, WI, 53706} 
 
\author{Matthew Kerr}
\affiliation{Space Science Division, Naval Research Laboratory, Washington, DC 20375}

\author{Roger Blandford}
\affiliation{
Kavli Institute for Particle Astrophysics and Cosmology, Stanford University, Stanford, CA 94305
}
\affiliation{
SLAC National Accelerator Laboratory, 2575 Sand Hill Road, Menlo Park, CA 94025
}

\author{Henrike Fleischhack}
\affiliation{%
 Department of Physics, Catholic University of America, Washington, DC 20064
}%
\affiliation{NASA Goddard Space Flight Center, Greenbelt, MD 20771}
\affiliation{Center for Research and Exploration in Space Science and Technology, NASA/GSFC, Greenbelt, MD 20771}

\author{Eric Charles}
\affiliation{
SLAC National Accelerator Laboratory, 2575 Sand Hill Road, Menlo Park, CA 94025
}%
\affiliation{
Kavli Institute for Particle Astrophysics and Cosmology, Stanford University, Stanford, CA 94305
}%

\date{\today}% It is always \today, today,
             %  but any date may be explicitly specified

\begin{abstract}
The existence of a ``knee'' at energy $\sim1\,{\rm PeV}$ in the cosmic-ray spectrum suggests the presence of Galactic PeV proton accelerators called ``PeVatrons". Supernova Remnant (SNR) G106.3+2.7 is a prime candidate for one of these. The recent detection \citep{Acciari_2009, HAWC_2020, Tibet_2021, LHAASO_2021} of Very High Energy (VHE; 0.1-100 TeV) gamma rays from G106.3+2.7 may be explained either by  the decay of neutral pions or inverse Compton scattering by relativistic electrons. We report an analysis of 12 years of {\it Fermi}-LAT \citep{2009ApJ...697.1071A} gamma-ray data which shows that the GeV--TeV gamma-ray spectrum is much harder and requires a different total electron energy than the radio and X-ray spectra, suggesting it has a distinct, hadronic origin. The non-detection of gamma rays below 10~GeV implies additional constraints on the relativistic electron spectrum. A hadronic interpretation of the observed gamma rays is strongly supported. This observation confirms the long-sought connection between Galactic PeVatrons and SNRs. Moreover, it suggests that G106.3+2.7 could be the brightest member of a new population of SNRs whose gamma-ray energy flux peaks at TeV energies. Such a population may contribute to the cosmic-ray knee and be revealed by future VHE gamma-ray detectors.
\end{abstract}

\maketitle

\begin{figure*}[t]
    \centering
    \includegraphics[width=.49\textwidth]{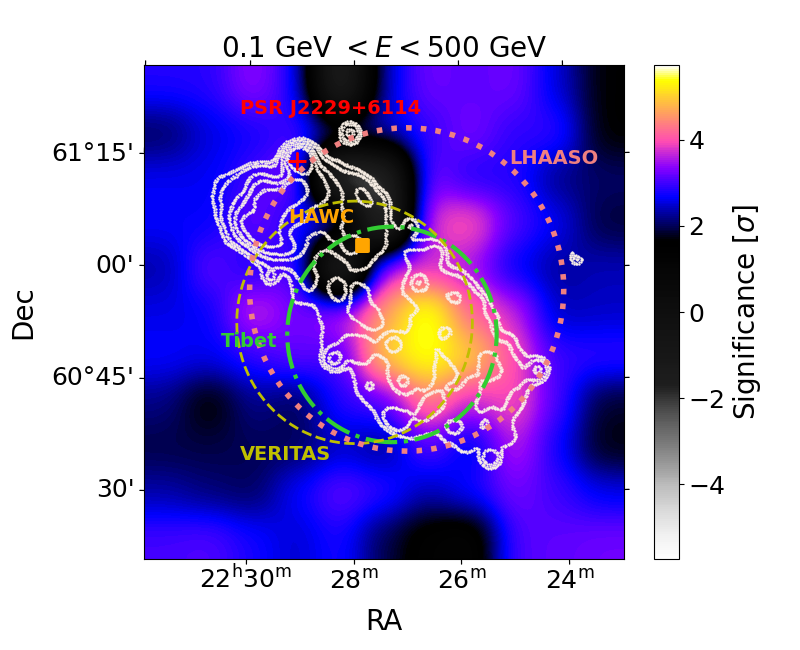}
    \includegraphics[width=.49\textwidth]{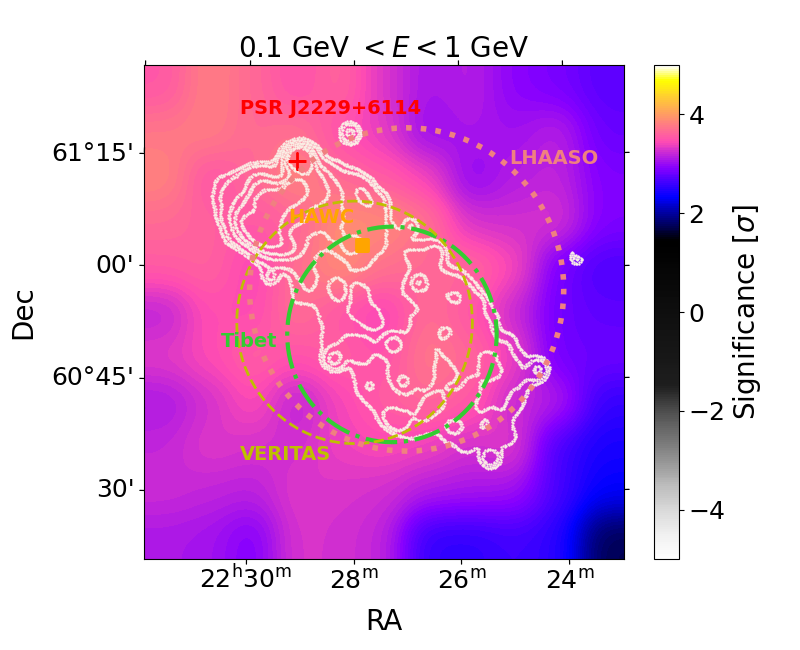}
    \includegraphics[width=.49\textwidth]{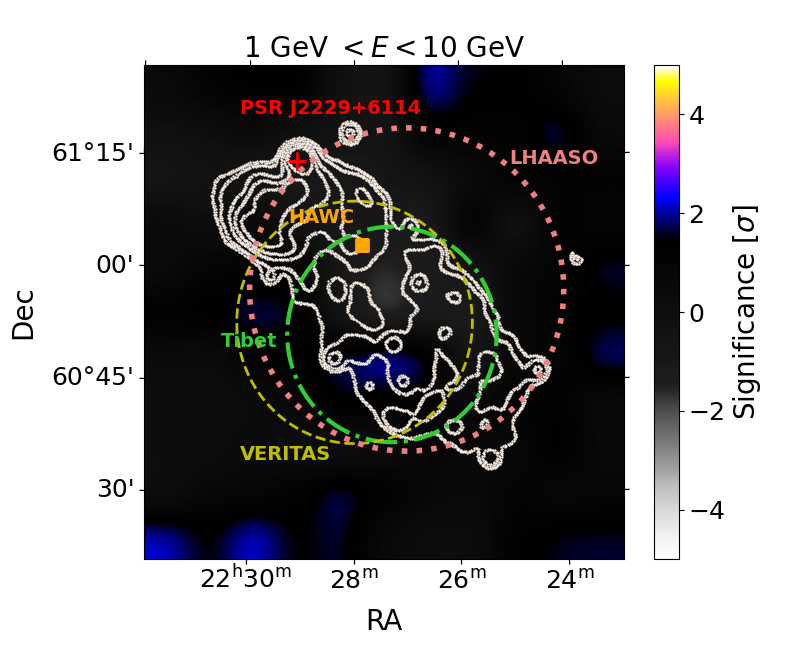}
    \includegraphics[width=.49\textwidth]{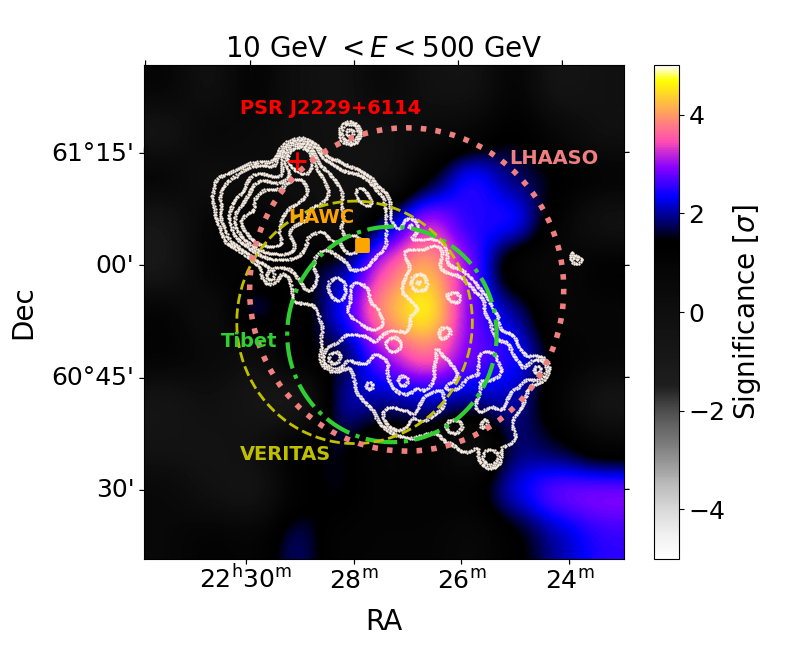}
    \caption{\label{fig:TSmap} Residual significance maps of the G106.3+2.7 region computed using the method of \citep{PSmapRef} from the analysis of 12 years of 0.1--500~GeV {\it Fermi}-LAT data (top left) and divided into three energy bins, 0.1--1~GeV (top right), 1--10~GeV (bottom left), and $>$10~GeV (bottom right). All maps except the $>$10~GeV one were computed using the weighted likelihood analysis \citep{3FGL} and phase-gated data (see the Supplementary Material), while the $>$10~GeV analysis used all of the data. The color scale indicates the statistical significance of a deviation between the data and the source model, evaluated on a grid with $0.1^\circ\times 0.1^\circ$ spacing. The maps are smoothed by Gaussian interpolation. For comparison, we show the radio continuum emission at 1420~MHz \citep{radio_2001} (white contours), the position of the pulsar PSR~J2227$+$6114 (red plus marker), the point source detected by HAWC \citep{HAWC_2020} (orange square marker), and extended gamma-ray emitting regions observed by VERITAS \citep{Acciari_2009} (yellow dashed circle), Tibet AS$\gamma$ \citep{Tibet_2021} (green dash-dotted circle), and LHAASO \citep{LHAASO_2021} (coral dotted circle).}
\end{figure*}

G106.3+2.7 is a comet-shaped, middle-aged ($\sim 10$~kyr) SNR at a distance of $\sim$800~pc \citep{Pineault_2000, radio_2001}. In the radio and X-ray bands, it is composed of a small ``head" structure in the north and an extended ``tail" in the southwest with lower surface brightness. The pulsar PSR~J2229+6114 and its wind nebula, the ``Boomerang" (with a length of about 3'), are located at the northern edge of the head (see Figure~\ref{fig:TSmap}) and are conjectured to result from the same supernova explosion that led to the formation of G106.3+2.7 \citep{radio_2001}. Non-thermal diffuse X-ray emission \citep{2020arXiv201211531G, Suzaku_2021} and radio emission~\citep{radio_2001} are detected from the entire SNR. The intensity in both bands increases toward PSR~J2229+6114 \citep{Pineault_2000,2020arXiv201211531G, Suzaku_2021}. The radio and X-ray spectra from XMM-Newton and Chandra are found to be harder in the head than in the tail \citep{Pineault_2000,2020arXiv201211531G}, though an analysis of the Suzaku data concludes that the photon index does not change with the distance from the pulsar \cite{Suzaku_2021}. 

The Very High Energy (VHE; 0.1-100 TeV) gamma-ray emission of the SNR appears to come from the tail \citep{Acciari_2009, HAWC_2020, Tibet_2021, LHAASO_2021}. The $68\%$ extension of the VHE emission is measured to be $0.23^\circ-0.45^\circ$ by different experiments, though the values are consistent within uncertainties.
The centroids of gamma-ray emission regions measured by VERITAS and Tibet and the best-fit position found by LHAASO overlap with a molecular cloud, while the VHE emission region in the HAWC data is consistent with both the pulsar and the molecular cloud due to the large position uncertainty.
When modeling the VHE counts rate spectrum (photons\,eV$^{-1}\,\rm cm^{-2}\,s^{-1}$) as a power law, $dN/dEdAdt \propto E^{-\alpha}$, the best-fit spectral index $\alpha$ is found between $2.3$ and $3.0$. 
The observed spectrum and the morphology may be explained by the interaction of hadronic cosmic rays and the molecular cloud, but a leptonic scenario, where gamma rays are produced by locally-accelerated relativistic electrons, is still possible \citep{Acciari_2009, HAWC_2020, Tibet_2021}. 

High-energy (0.1-100~GeV) gamma-ray observations, especially below 10 GeV, are crucial to breaking the degeneracy of the hadronic and leptonic scenarios. A previous analysis \citep{Xin_2019} using 10 years of {\it Fermi}-LAT data above 3~GeV found an excess in the tail with a test statistic (TS) \footnote{The Test Statistic is defined as $\mathrm{TS}=-2\log({\cal{L}_{\rm max,0}}/{\cal{L}}_{\rm max, 1})$, where $\cal{L}_{\rm max, 0, 1}$ are maximum likelihood values for models without and with an additional source.} of $35.5$ and a disk morphology of radius $0.25^\circ$, while the properties of G106.3+2.7 below 3~GeV remained unexplored. Such low-energy analysis is complicated because PSR~J2229+6114, also known as 4FGL~J2229.0+6114 in the fourth {\it Fermi}-LAT catalog of gamma-ray sources (4FGL-DR2 \citep{4FGL, 4FGLDR2}), dominates the gamma-ray emission of the entire region up to a few GeV as explained in the Supplementary Material.

{\it Results of the Fermi-LAT analysis.---}We searched for high-energy gamma-ray signals using 12 years of {\it Fermi}-LAT data selected to include only rotational phases when the gamma-ray emission of the pulsar is minimal to avoid this background contamination. By eliminating 50\% of the observing time, we reduce the background from PSR~J2229$+$6114 by $>$95\% at low energies (0.1~GeV) and by 99\% above 1~GeV (details are provided in Figures~\ref{fig:allCounts} and \ref{fig:pulsar} in the Supplementary Material). 
Figure 1 presents the maps of the significance of the deviations between the LAT data and the source model, comprising sources in the 4FGL-DR2 catalog \citep{4FGLDR2} and diffuse backgrounds, in the full energy range and three energy bins, $0.1-1$~GeV, $1-10$~GeV, and above 10~GeV. The deviation significance is computed according to \citep{PSmapRef}, applying for each pixel an energy-dependent spatial selection that roughly follows the LAT point-spread function (PSF). Such maps allow us to detect potential excess emissions (point-like or with a relatively small, degree-scale extension, as explained in the Supplementary Material), whose spatial and spectral characteristics are subsequently investigated with a more detailed analysis as explained below.
%{\st{To reduce the number of free parameters in the search the source spectrum is assumed to be a power-law with a fixed index $E^{-2}$, because many sources have similar spectral shapes.}}  }  
 %in a grid with $0.1^\circ\times 0.1^\circ$ spacing, 
In the lowest-energy bin, excess emission is present in the entire vicinity of the  remnant with low significance. As the 68\% containment radius of the PSF of the LAT below 1~GeV is larger than $2^\circ$, the photons may also come from nearby sources or the Galactic plane. In the intermediate energy bin, no significant excess or deficit is observed inside the remnant. A source is clearly present in the highest-energy bin, with the best-fit position consistent with the VHE gamma-ray emitting site. 

Above 10~GeV, the pulsar emission is negligible, and the 68\% containment radius of the PSF of the LAT ($\lesssim 0.2^\circ$) is narrower than the angular distance between the pulsar and the gamma-ray emitting site, so we use all the data for the analysis. When fitting the data with a point-source morphology and a power-law energy distribution, we obtain ${\rm TS}=36.5$ with four free parameters, including the coordinates of the source position, and the flux normalization and spectral index. This corresponds to $5.2\,\sigma$ standard deviations. The best-fit spectral index is $1.72\pm 0.20$  and the maximum likelihood coordinates are  $({\rm RA, Dec})=( 336.71^\circ\pm 0.03^\circ, 60.90^\circ\pm0.03^\circ)$ (J2000), corresponding to Galactic coordinates $(l, b) = (106.24^\circ \pm  0.03^\circ, 2.81^\circ \pm 0.03^\circ)$. The position and the differential energy flux of the emission are consistent with those found in the TeV measurements. We also fit the data with extended spatial profiles and summarize the results in the Supplementary Material. In general an extended morphology yields a larger TS since the extended model has one more degree of freedom than the point-source model. The most favored extended model with a Gaussian radial profile yields $\Delta {\rm TS} = 7.4$ with $0.2^\circ$ radius, which is not a significant improvement over the point-source model ($<3\sigma$).  We thus conclude that gamma-ray emission from G106.3+2.7 is unresolved in the LAT data above 10~GeV.

\begin{figure*} 
    \centering
   \includegraphics[width=.75\textwidth]{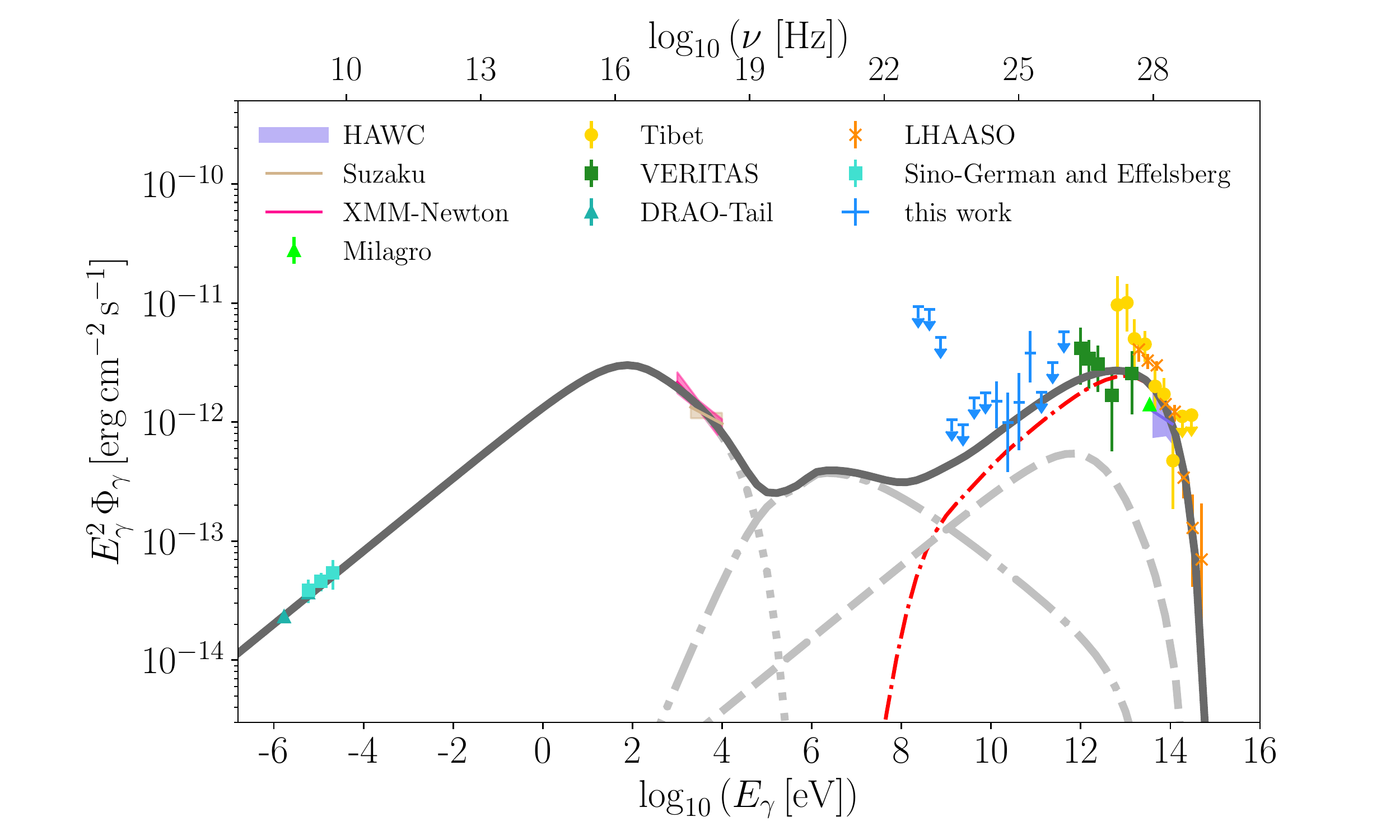}
    \caption{\label{fig:sed} Broadband spectral energy distribution of supernova remnant G106.3+2.7. The multi-wavelength data include radio \citep{Pineault_2000, 2011A&A...529A.159G},  X-ray \citep{2020arXiv201211531G, Suzaku_2021},  and
VHE gamma-ray \citep{Acciari_2009, HAWC_2020, Tibet_2021, LHAASO_2021}.  Error bars indicate $1\,\sigma$ uncertainties. For the LAT data points, 95\% upper limits are shown when $\rm TS <4$, otherwise $1\,\sigma$ error bars are shown. The VERITAS flux points are scaled up by a factor of 1.62 from the original values to account for the gamma-ray signals outside the signal extraction region of the analysis \citep{Acciari_2009, Tibet_2021}. For comparison, the multiwavelength spectrum from a hybrid model including an electron population (in grey color) and a proton population (in red color) is shown. The injection spectra of both populations are assumed to be exponentially cut off power laws (refer to the text for the best-fit values of the spectral parameters). The electrons produce radio to X-ray photons through synchrotron emission in a magnetic field (dotted curve), hard X-ray to sub-GeV gamma-ray through Bremsstrahlung emission with gas in the interstellar medium (dash dotted curve), and gamma rays above 10~GeV through inverse Compton scattering of the CMB (dashed curve). The protons produce gamma rays through gas interaction (dash-dotted curve).
    } 
\end{figure*}

We then use these $>$10~GeV results to guide the analysis above 0.1~GeV, where the angular resolution is poorer and the diffuse background emission is larger. To reduce the impact of diffuse emission below $\sim 3$ GeV, we maximize a likelihood function that includes de-weighting the photons in that energy range \citep{3FGL}. 
Using the phase-gated data, fixing the G106.3+2.7 position to the best-fit values found in the $>$10~GeV analysis, and leaving the spectral parameters of the SNR free, we find a TS of $20.8$ with a point-source morphology and $34.8$ with a 2D Gaussian template with 68\% containment radius of $0.2^\circ$, corresponding to $4.2\,\sigma$ and $5.6\,\sigma$ standard deviations, respectively.
Below we take the extended template as a benchmark model which provides the greatest likelihood of the SNR among the models that we studied. 
The blue data points in Figure~\ref{fig:sed} show the spectral energy distribution (SED) of the SNR from the fit results from the benchmark model. The SED is calculated by binning the photons into four bins per decade in energy and performing a weighted likelihood analysis in each energy bin. For all spatial templates we tested, no significant emission is detected below 10~GeV.

Since the SNR is not bright, the measured flux may be affected by the modeling of nearby faint emission regions. We discuss this impact in the Supplementary Material, but note that transferring more of the GeV emission from the SNR to any background source than in the benchmark model only strengthens the evidence, discussed below, for the presence of protons accelerated by SNR G106.3 +2.7.

{\it Multi-wavelength observation and broadband SED.---}We combine the {\it Fermi}-LAT spectral results with the radio, X-ray, and VHE observations of the remnant and use the broadband SED to constrain physical models through the Markov chain Monte Carlo (MCMC) technique. Nested models are then compared using both the likelihood-ratio test and the Bayesian information criterion (BIC) \citep{10.1214/aos/1176344136}. Since Wilks' theorem \citep{10.1214/aoms/1177732360} applies only to nested models, our comparisons of other models are based only on BIC values.

Physical models of VHE gamma-ray emission invoke high-energy leptons, hadrons, or a combination of the two. In the one-component, leptonic version, a population of relativistic electrons is continuously injected by either the magnetosphere or the nebula of the pulsar, or the supernova remnant shock front. The electrons are  confined by magnetic turbulence over the source age $t_{\rm age}\sim 10$~kyr and cool through synchrotron, bremsstrahlung, and inverse Compton radiation. The electron spectrum can be modeled by a power-law spectrum with an exponential cutoff, ${dN_e}/{dE} = N_{e0}\,\gamma^{-\alpha_e}\,\exp(-E / E_{e\,\rm max})$, where $\gamma = E / m_e c^2$ is the Lorentz factor. To calculate the flux of the bremsstrahlung radiation, we adopt a gas density equal to the average interstellar medium density, $n_{\rm gas}=1\,\rm cm^{-3}$. This mean density is supported by the presence of H\,{\sc i} and CO gas associated with the SNR \citep{radio_2001}, and $n_{\rm gas}$ would be much greater at the location of the molecular cloud \citep{Heyer_1998, Tibet_2021}. The value is consistent with the expected SNR gas density in the Taylor-Sedov phase \cite{1959sdmm.book.....S, 2008ARA&A..46...89R}, $n_{\rm Sedov} \sim 2.1\,(E_{\rm exp}/10^{51}\,\rm erg)\,(t_{\rm age}/10\,\rm kyr)^2(R/10\,\rm pc)^{-5}\,\rm cm^{-3}$, where $E_{\rm exp}$ is the total energy released by the supernova explosion and $R$ is the size of the SNR.   
As no conspicuous infrared emission is found from the SNR \citep{radio_2001}, the inverse Compton radiation is calculated using the Cosmic Microwave Background (CMB) and a background Galactic far infrared (FIR) emission with temperature 30~K and  energy density $0.3\,\rm eV\,cm^{-3}$ \cite{2016PhRvD..94f3009V} as the target radiation background. The electron spectrum is obtained by numerically solving the transport equation assuming a continuous injection as described in the Supplementary Material. The top panel of Figure~\ref{fig:sed_supp} presents the best-fit one-component leptonic model. We find that the leptonic model fails to explain the multi-wavelength emission for two main reasons. First, the radio and non-thermal X-ray emission suggest an electron spectral index $\alpha_e = 2.42_{-0.06}^{+0.04}$, whereas a harder electron spectrum with  $\alpha_e = 2.14_{-0.11}^{+0.09} $ is needed to explain the gamma-ray flux from sub-GeV to a few TeV. The best-fit electron spectrum based on the radio to X-ray data has a total energy  $W_e
\sim 10$ times higher than the $W_e$ derived from the gamma-ray data. The differences in $\alpha_e$ and $W_e$ are smaller in a less physical model where the cooling of electrons is not included. Second, the Bremsstrahlung emission by low-energy and high-energy electrons appear at the energy ranges where the end of the synchrotron emission and the beginning of the inverse Compton emission also contribute. Together these components are in tension with the X-ray measurements at 2-10~keV and the 1-10~GeV {\it Fermi}-LAT upper limits, respectively. As a result, the best-fit model flux is 5--10 times lower than the measured flux above 100~TeV. The tension would be stronger if $n_{\rm gas}$ is higher than $1\,\rm cm^{-3}$.

By contrast, a similarly simple hybrid model that includes a hadronic contribution naturally accounts for these spectral features. The proton spectrum can be modeled by a single power-law spectrum with an exponential cutoff, ${dN_p}/{dE} = N_{p0}\,\gamma^{-\alpha_p}\,\exp(-E / E_{p\,\rm max})$, where $\gamma = E/m_p c^2$ is the Lorentz factor of protons. The relativistic protons interact with gas in the surrounding medium and produce neutral pions that decay into gamma rays. Figure~\ref{fig:sed} shows the best-fit model with the following parameters, $\log (N_{p0} / \rm eV^{-1})=40.27_{-0.77}^{+0.50}$, $\alpha_p = 1.73_{-0.16}^{+0.12}$, $\log (E_{p,\rm max}/\rm eV)= 14.95_{-0.13}^{+0.13}$, $\log (N_{e0} / \rm eV^{-1}) = 47.69_{-0.53}^{+0.57}$, $\alpha_e = 2.39_{-0.06}^{+0.05}$, $\log (E_{e,\rm max}/\rm eV) = 14.54_{-0.15}^{+0.26}$, and $B=8.99_{-3.54}^{+4.85}\,\mu$G.  
By integrating the energy flux $E dN/dE$ above the rest masses of electrons and protons, we find a total proton energy $W_p = 3.3\times 10^{48}\,\rm erg$ and electron injection energy  $W_e=5.3\times10^{47}$~erg, respectively. The required particle acceleration efficiency, $\epsilon_{\rm CR}\sim (W_e+W_p) / E_{\rm k} = 0.4\%\,(E_{\rm k}/10^{51}\,\rm erg)^{-1}$ with $E_{\rm k}$ being the kinetic energy of the SNR, can be achieved by a typical SNR. The ratio of the proton and electron acceleration efficiencies is $W_p/W_e\sim 10$, which is consistent with the finding from individual SNR observation and collectively in the cosmic-ray spectrum that SNRs accelerate protons $\sim 10-100$ times more efficiently than electrons \citep{2013A&ARv..21...70B}.  
The hybrid model is significantly preferred over the  one-component leptonic model by $\Delta{\rm BIC} =-20.6$, and $\rm TS = 31.7$ which corresponds to $5.0\,\sigma$.

We further investigate whether a more complicated leptonic model could better explain the data. Figure~\ref{fig:sed_supp} shows such a model where two populations of electrons are invoked to resolve the difficulty of the one-component model in explaining the radio to X-ray and gamma-ray measurements simultaneously. The two components are assumed to be accelerated by different mechanisms, such as the pulsar and remnant shocks, and thus have different spectra. The two populations of electrons are assumed to both contribute to the tail region. This model is, however, again disfavored by data. The lepto-hadronic hybrid model yields a much lower BIC, $\Delta \rm BIC = -20.1$,  compared to the two-component leptonic model, suggesting that it is significantly preferred. Moreover, as the radio and X-ray intensities are observed to span the entire remnant with their fluxes increasing toward the pulsar in a similar way, they likely share the same production mechanism. The two-component leptonic model would break down the natural connection of the two bands.

Besides poorer fits to data, leptonic models share a common weakness. If the X-ray and gamma-ray fluxes come from the synchrotron and inverse-Compton emission of electrons, respectively, the magnetic energy density of the SNR would be comparable to the energy density of the radiation field. Since gamma rays above 10~TeV are mostly up-scattered CMB photons, the field strength cannot be much higher than $\sim 3\,\mu$G. Such a field strength is too low for the acceleration of near-PeV electrons, which are needed to produce 100--500~TeV gamma rays. The maximum electron energy that can be accelerated by the remnant shocks is \cite{1987PhR...154....1B} $E_{e,\rm max}= 188\,{\rm TeV}\, \eta^{1/2}\,(B/3\,\mu{\rm G})^{-1/2}\,(v_{\rm sh}/3000\,\rm km\,s^{-1})$, where  $v_{\rm sh}$ is the shock velocity and $\eta=\delta B^2/B^2$ is the degree of magnetic field fluctuations that characterizes the acceleration efficiency. Such a low field supports neither an electron acceleration by the pulsar or its nebula.  Because in that case, as the physical condition of the gamma-ray emitting site is not much different from the average ISM condition \citep{2001SSRv...99..243B}, GeV gamma rays should have been found in the head region of G106.3+2.7.

{\it Summary and discussion.---}SNRs have long been proposed as efficient accelerators of cosmic rays  \citep{2008ARA&A..46...89R, 2011JCAP...05..026C} up to PeV energies \citep{2013MNRAS.431..415B}. However, only a handful \citep{2018A&A...612A...3H, 2019ICRC...36..674F}, 
%$\sim 25$ \footnote{http://tevcat.uchicago.edu}, 
out of hundreds of radio-emitting SNRs, have been observed to emit VHE radiation with a hard spectrum. 
The scarcity of PeVatron candidates and the rareness of SNRs with VHE emission make SNR~G106.3+2.7 a unique source. Our study provides strong evidence for proton acceleration in this nearby SNR, and by extension, supports a potential role for G106.3+2.7-like SNRs 
in meeting the challenge of accounting for the observed cosmic-ray knee using Galactic sources. 
%{\ke {\st {A small fraction of the SNR population, $\sim 6\% \,\left(E_{\rm k}/10^{51}\,\rm erg\right)^{-1}\left(\epsilon_{\rm CR}/10\%\right)^{-1}\left(r_{\rm SN}/1.7\times10^{-2}\,\rm yr\right)^{-1}$ where $r_{\rm SN}$ is the birth rate of supernovae in the Galaxy, would be sufficient to explain the intensity of Galactic cosmic rays above 100~TeV, }}}  
Future VHE gamma-ray observatories such as Cherenkov Telescope Array \citep{2019scta.book.....C} and Southern Wide-field Gamma-ray Observatory \citep{2019BAAS...51g.109H} could reveal the subgroup of SNRs which has gamma-ray energy flux peaked at TeV energies like G106.3+2.7. 

\vspace{2em}

\begin{acknowledgments}
The \textit{Fermi}-LAT Collaboration acknowledges support for LAT development, operation and data analysis from NASA and DOE (United States), CEA/Irfu and IN2P3/CNRS (France), ASI and INFN (Italy), MEXT, KEK, and JAXA (Japan), and the K.A.~Wallenberg Foundation, the Swedish Research Council and the National Space Board (Sweden). Science analysis support in the operations phase from INAF (Italy) and CNES (France) is also gratefully acknowledged. This work performed in part under DOE Contract DE-AC02-76SF00515. 

We thank Seth Digel for helpful comments on the manuscript. We thank Philippe Bruel for useful suggestions regarding the usage of PS maps. 
The work of K.F. is supported by the Office of the Vice Chancellor for Research and Graduate Education at the University of Wisconsin-Madison with funding from the Wisconsin Alumni Research Foundation. K.F. acknowledges support from NASA through the Fermi Guest Investigator Program (NNH19ZDA001N-FERMI, NNH20ZDA001N-FERMI) and from National Science Foundation (PHY-2110821). Work at NRL is supported by NASA. H.F. acknowledges support by NASA under award number 80GSFC21M0002.

\end{acknowledgments}

\newcommand {\aap}      {A\&A}
\newcommand {\apjl}     {ApJ Letters}
\newcommand {\apjs}     {The Astrophysical Journal Supplement Series}
\newcommand {\araa}     {Annual Review of Astronomy and Astrophysics}
\newcommand {\aapr}     {A\&A Reviews}
\newcommand {\jcap}     {JCAP}
\newcommand {\physrep}  {Physics Reports}
\newcommand {\mnras}    {MNRAS}
\newcommand {\ssr}      {Space Science Reviews}

\bibliography{references}% Produces the bibliography via BibTeX.

\clearpage 
\appendix
\section{Appendices}

\section{Fermi-LAT Data Analysis}\label{appendix:latAnalysis}
We analyze $12$ years of Pass 8 data \citep{2013arXiv1303.3514A, 2018arXiv181011394B} %\footnote{https://fermi.gsfc.nasa.gov/ssc/data/analysis/documentation/Cicerone/Cicerone\_Data/LAT\_DP.html} 
collected from August 04, 2008, to August 04, 2020, and select gamma-ray events with energy between $100$~MeV and 500~GeV. 
We apply the {\texttt {P8R3\_SOURCE}} event selection and the corresponding {\texttt {P8R3\_SOURCE\_V3}} LAT instrument response functions. The energy resolution of the Pass~8 data is $<$10\% above 1~GeV and worsens to $\sim$20\% at 100~MeV. It also worsens again above $\sim 1$~TeV when much of the particle shower energy escapes the calorimeter. The region-of-interest (ROI) of our analysis is defined as the $15^\circ \times 15^\circ$ region surrounding the supernova remnant.  
The baseline model of the ROI includes known sources in the 4FGL-DR2 catalog together with the corresponding model of diffuse interstellar emission {\texttt{gll\_iem\_v07.fits}} \footnote{\url{https://fermi.gsfc.nasa.gov/ssc/data/access/lat/BackgroundModels.html}} and the isotropic diffuse model. At the beginning of the analysis, we optimize the baseline model of the ROI using the \texttt {analyze-roi} function of \texttt {fermipy}, which performs sequential likelihood fits to obtain best-fit spectra and locations for the 4FGL-DR2 sources in the ROI. We adopt a binned likelihood function as in Ref.~\cite{1996ApJ...461..396M}.

To determine a timing solution valid over the time span of the data, we used a maximum likelihood technique operating directly on the photons.  Using the 4FGL-DR2 sky model, we computed photon weights $w_i$ using {\texttt{gtsrcprob}}.  For a timing model with parameters $\mathbf{\lambda}$, the pulse phase of a photon received at time $t_i$ is denoted $\phi(t_i,\mathbf{\lambda})$, and the pulse profile is $f(\phi)$, normalized such that $\int_0^1 f(\phi)\,d\phi=1$.  Then, the log likelihood describing the data is
\begin{equation}
\log\mathcal{L} = \sum_i \log (w_i f (\phi(t_i,\mathbf{\lambda}) +(1-w_i)),
\end{equation}
and estimators for the timing model parameters $\mathbf{\lambda}$ can be obtained by maximizing the log likelihood.  We estimate $f(\phi)$ analytically as the sum of three wrapped gaussians.

PSR~J2229$+$6114 exhibits strong timing noise, or random wander of its spin phase, which we have modeled as a stationary random process with a power spectral density that follows a power law \citep{Kerr15c}.  We approximate the timing noise process as a truncated Fourier series, with the amplitudes of the coefficients constrained to follow the power-law power spectral density. One-hundred ten coefficients are required to reach the intrinsic ``white noise'' level of the data.  To evaluate $\phi(t_i,\lambda)$, we use \textsc{PINT} \citep{Luo21}.The resulting ephemeris is made available on the Fermi Science Support Center pulsar ephemeris web page \footnote{\url{https://fermi.gsfc.nasa.gov/ssc/data/access/lat/ephems/}}. 

\begin{figure*}[t]
    \centering
    \includegraphics[width=.49\textwidth]{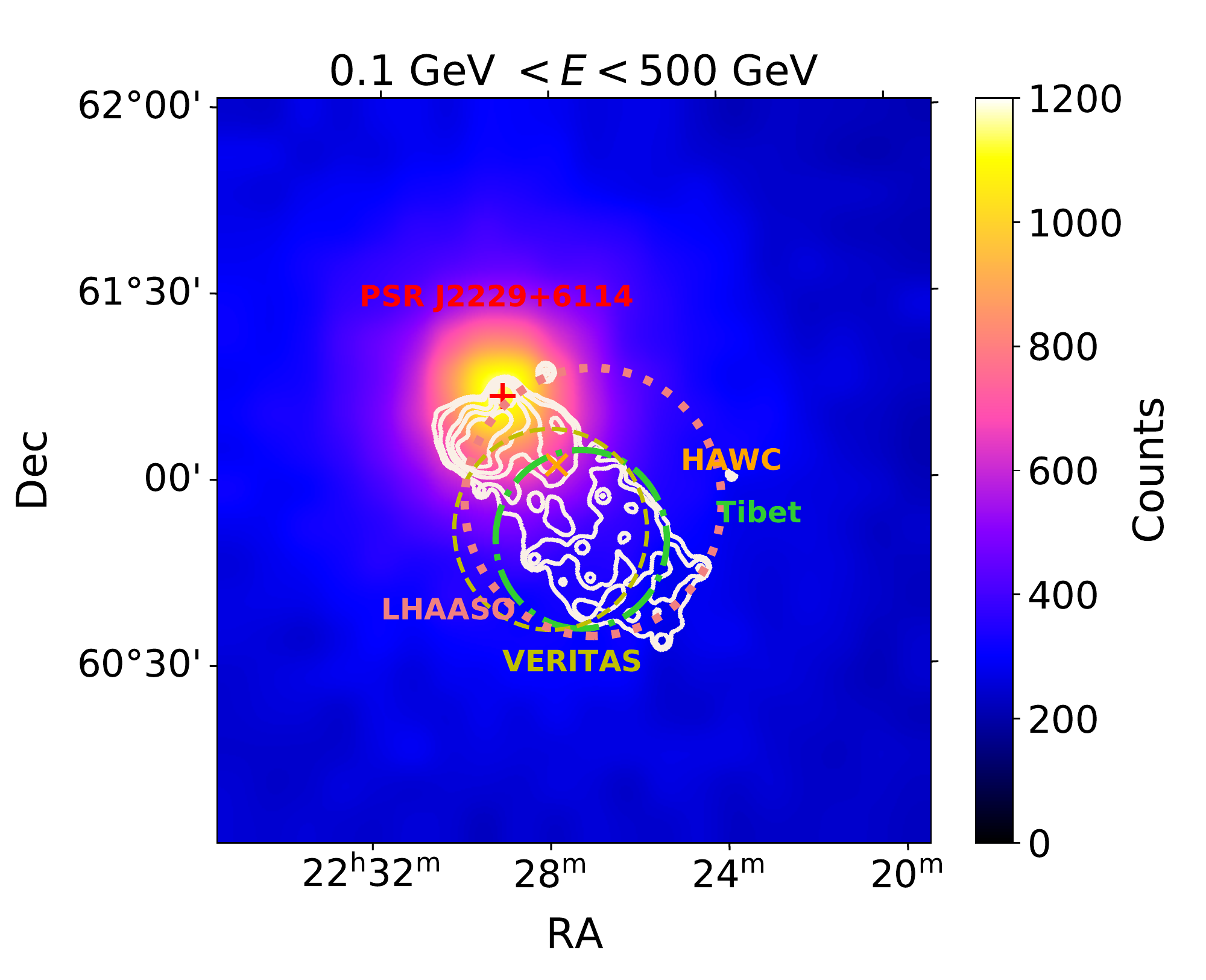}
    \includegraphics[width=.49\textwidth]{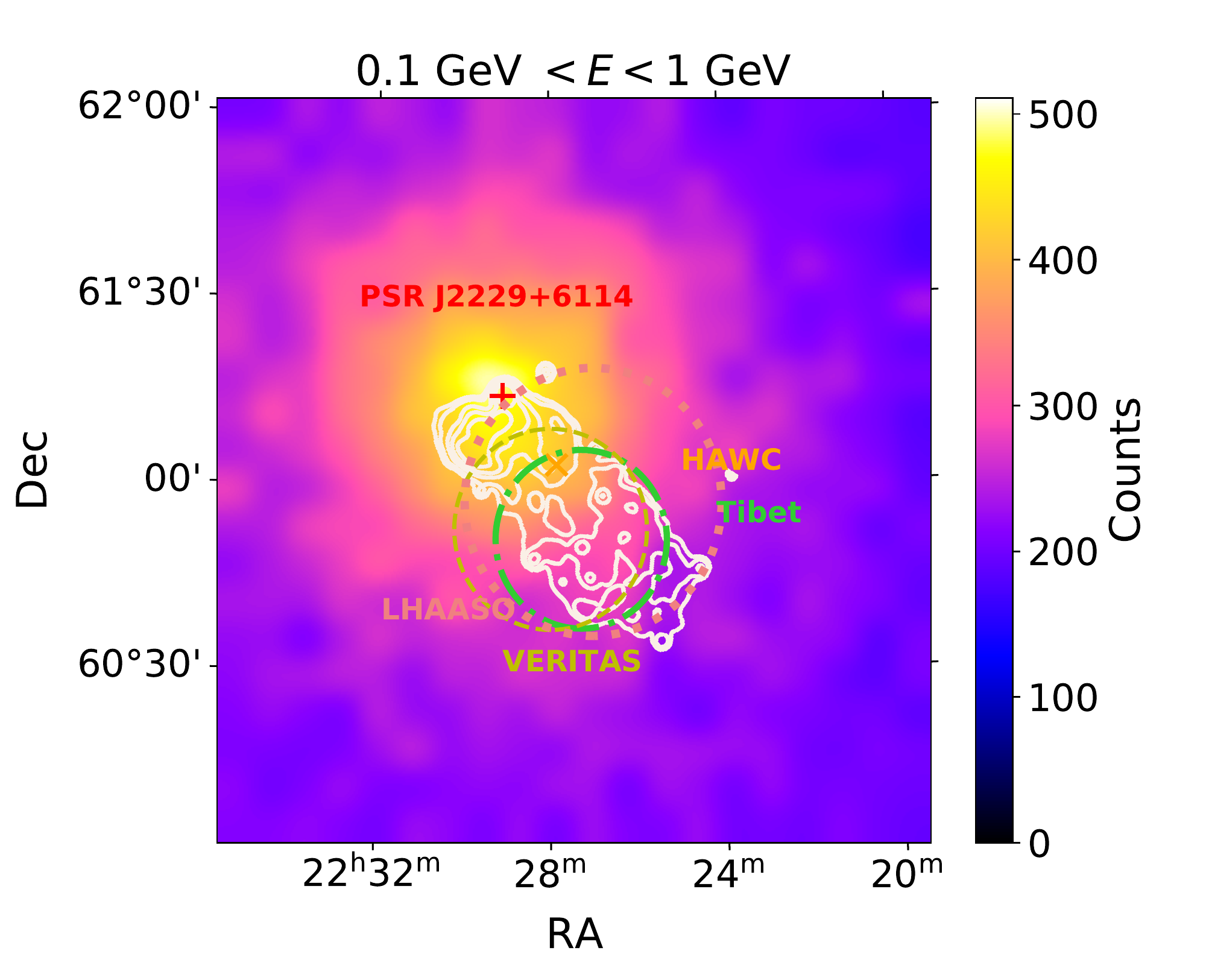}
    \includegraphics[width=.49\textwidth]{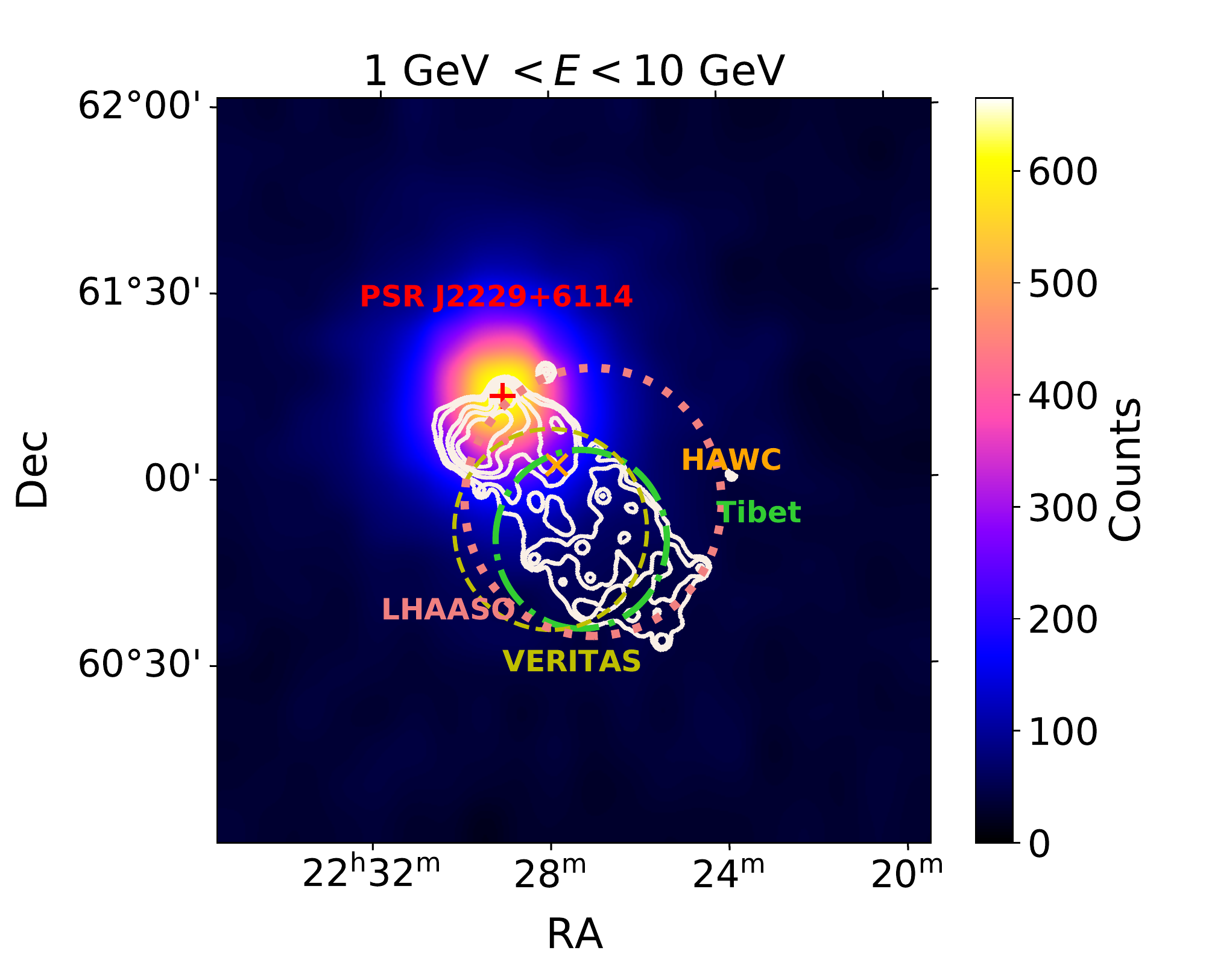}
    \includegraphics[width=.49\textwidth]{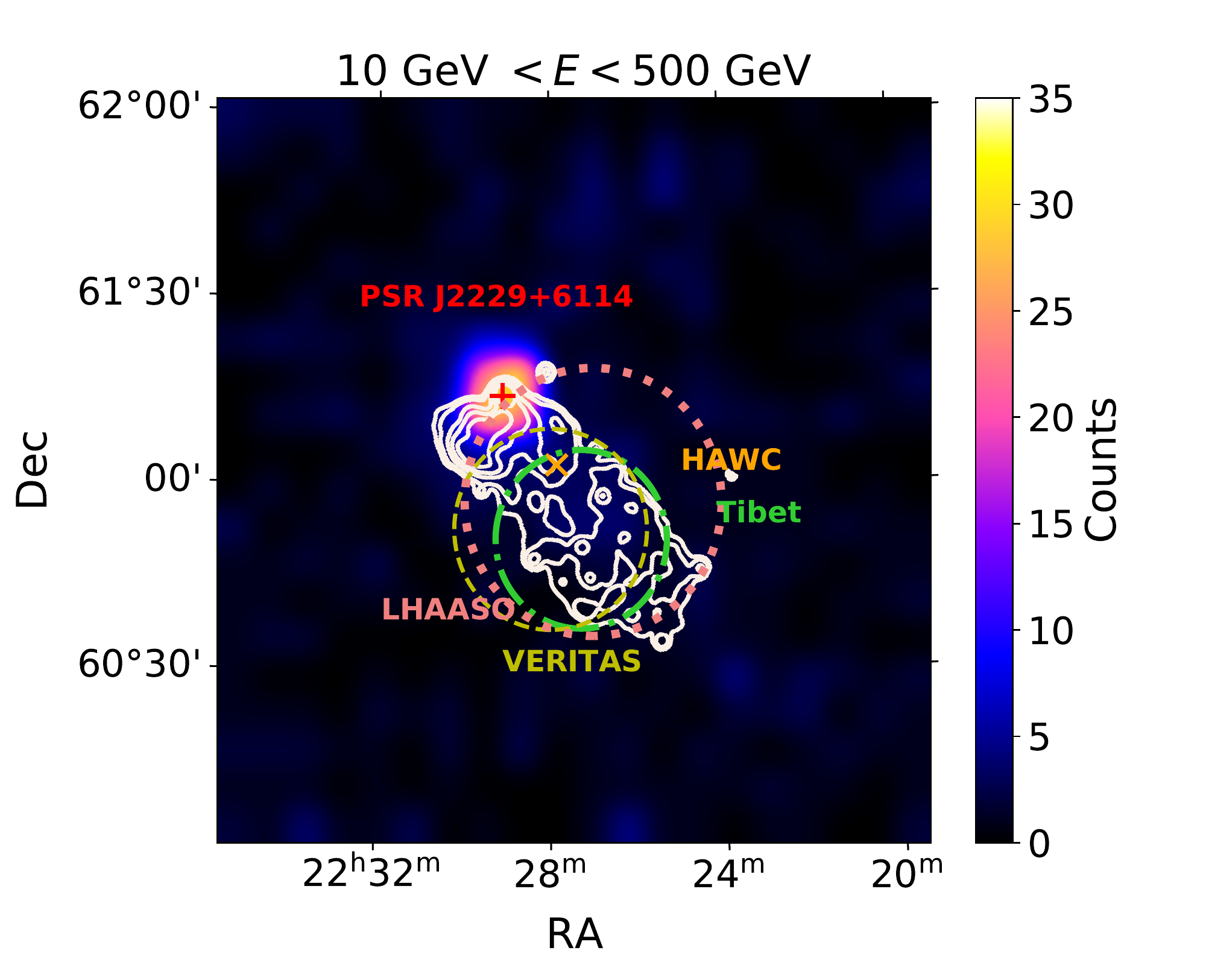}
    \caption{\label{fig:allCounts} Photon counts maps between 100~MeV and 500~GeV from the 12-year {\it Fermi}-LAT observation (top left), and divided into three energy bins, 0.1--1~GeV (top right), 1--10~GeV (bottom left), and $>$10~GeV (bottom right).  The maps are smoothed by Gaussian interpolation. The red cross marker indicates the position of the pulsar. The yellow, orange, and green markers indicate the VHE gamma-ray sources detected by  VERITAS \citep{Acciari_2009}, HAWC \citep{HAWC_2020}, Tibet AS$\gamma$ \citep{Tibet_2021}, and LHAASO \citep{LHAASO_2021}, respectively. PSR~J2229+6114 contaminates the G106.3+2.7 vicinity in all energy bins.  }
\end{figure*}

\begin{figure}[t]
    \centering
    \includegraphics[width=.46\textwidth]{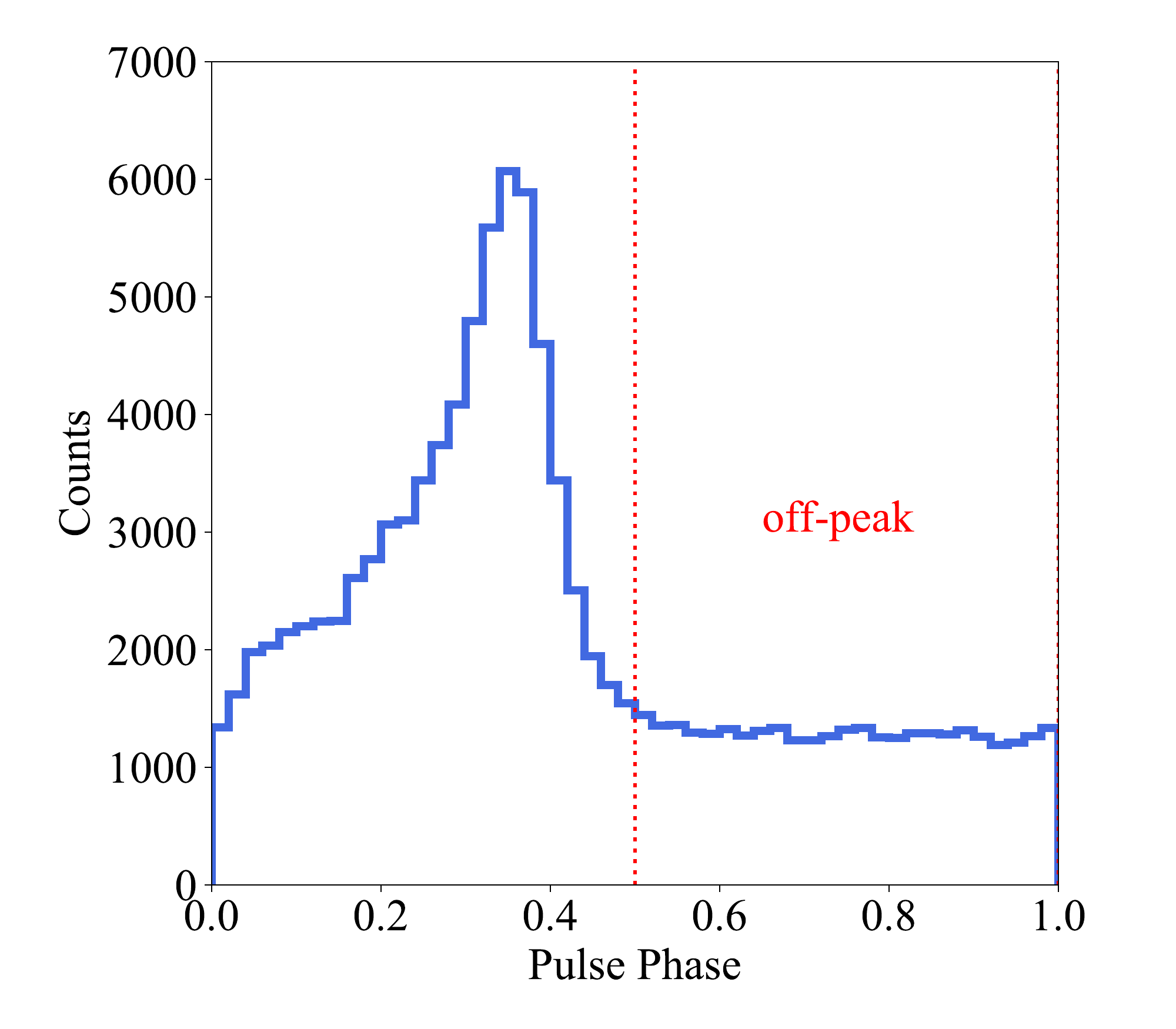}
    \caption{\label{fig:pulsar} Pulse profile of PSR~J2229+6114 between 100~MeV and 10~GeV including photons within $1^\circ$ of the pulsar. The red dotted lines at pulse phases $\phi=0.5$ and $\phi=1.0$ show the off-peak interval, which is used for the unpulsed data analysis in this work. }
\end{figure}

Figure~\ref{fig:allCounts} presents the 0.1-500~GeV counts map of the sky region surrounding G106.3+2.7. It suggests that the emission of the entire region is dominated by the pulsar PSR~J2229+6114 over the entire energy range. Figure~\ref{fig:pulsar} shows the pulse profile of the pulsar, which is computed using photon counts within $1^\circ$ of the pulsar between 0.1 and 10~GeV. Since the light curve of the pulsar is highly peaked, the gating based on photon phase is very efficient at suppressing the signal from the pulsar. A fraction 97.3\% (99.3\%) of the flux of the pulsar above 0.1~GeV (1~GeV) is in the pulsed component. The main pulse is almost entirely contained in the phase range [0-0.5], so by eliminating only half of the data we significantly reduce the signal from PSR~J2229$+$6114, making characterization of emission from the SNR much more robust against any mismodeling of the residual pulsar emission. For this analysis, we use data taken during pulsar phase range [0.5-1], which is shown as the off-peak phase in Figure~\ref{fig:pulsar}.

The significance of the deviation between the LAT data and the source model (see Figure~\ref{fig:TSmap}) is computed using the PS with the {\texttt{gtpsmap}} script \footnote{ \url{https://fermi.gsfc.nasa.gov/ssc/data/analysis/user/} } with default values of the parameters \texttt{psfpar0} $=4.0^\circ$, \texttt{psfpar1} $=100$~MeV, \texttt{psfpar2} $=0.9$, and \texttt{psfpar3} $=0.1^\circ$. PS is an estimator for the significance of the deviation between data and model, defined as $|{\rm PS}|=-\log_{10}($p-value$)$ \cite{PSmapRef}. The p-value is converted to a number of standard deviations using a two-sided Gaussian distribution.  The sign of the PS is determined by the sum of the residuals for all energy bins in sigma units. A PS map is similar to a test statistic (TS) map but may have negative values for cases of over-prediction of the data. The PS computation does not assume any spectral shape. Without a priori of the nature of the potential deviations, we usually use the point-source optimized parameters, which implies a moderate loss of sensitivity for extended deviations. A deviation due to a point source creates a peak in the PS map when using the point-source optimized parameters. If, with these parameters, the PS map has a rather flat maximum, it is likely that the deviation is due to an extended source. The PS maps are a qualitative, first step in the analysis. For the actual modeling of the supernova remnant, we model both the source morphology and spectral shape more precisely.

The systematic uncertainties in this {\it Fermi}-LAT analysis mainly arise from the background diffuse emission models, including both the Galactic and isotropic emission models. We employ a weighted likelihood analysis \cite{3FGL} (with the systematic uncertainty of the interstellar emission models set to $\epsilon = 3\%$ and using the data weights) to address the uncertainty. The weighted likelihood analysis is a conservative procedure to account for the systematic uncertainty by increasing the width of the log likelihood distribution from each energy band according to how sensitive it is to the diffuse contribution. This is done by reducing the weighting of the likelihood whenever the statistical precision on the Galactic diffuse emission is better than the systematic variations of the diffuse models.
The uncertainties from LAT analyses in this work include both statistical and systematic uncertainties. We neglect the residual systematic errors from the instrument response functions because they are small compared to the statistical uncertainties. 

\section{Analysis above 10~GeV} \label{appendix:10GeV}

\begin{table*}
\caption{Positions and spectral indices derived from fits to the 12-year {\it Fermi}-LAT data above 10~GeV   \label{tab:1s_10GeV}}
\begin{ruledtabular}
\begin{tabular}{ccccc}
  Morphology  &  Radius  & Best-fit Position  & Best-fit Index  &  TS     \\ 
  &  [degree]  &  (RA, Dec) [degree]  &   &    \\
  \hline
 Point Source    \space    & None  & (336.71, 60.90) & 1.72     &  36.5  \\
   Radial Gaussian  \space     & 0.15   & (336.67, 60.90) & 1.85  &  43.0  \\
   Radial Gaussian  \space    & 0.2   & (336.65, 60.91)  & 1.92 &  43.8  \\
   Radial Gaussian  \space     & 0.25    & (336.63, 60.90) &  1.98 & 43.8  \\
   Radial Gaussian  \space      & 0.3  & (336.60, 60.89) &  2.02 & 43.5  \\
   Radial Disk  \space     & 0.2  & (336.60, 60.92)   & 1.92 & 41.3 \\
\end{tabular}
\end{ruledtabular}
\end{table*}

\begin{table*}
 \caption{Results of fits to the unpulsed {\it Fermi}-LAT data at 0.1--500~GeV with the weighted likelihood \label{tab:BG1}}  
\begin{ruledtabular}
\begin{tabular}{ccccccccc} 
 Source & Morphology &  Radius & RA\footnotemark[1]  & Dec\footnotemark[1] & TS  & Index\footnotemark[1] &  Prefactor\footnotemark[2] & $\Delta \rm TS_{ROI}$  \\
 &  & [degree]  &  [degree] &   [degree]  &   &   &  & \\
\hline
Model A (one new source) & & & & & & & & \\
\hline
G106.3+2.7 &  Point Source    \space  & None    &336.71$^*$   & 60.90$^*$ &  20.8 & 1.72$^*$ & 1.43 & 20.5  \\
\hline
G106.3+2.7 &  Radial Gaussian   \space  & 0.2    &336.65$^*$  & 60.91$^*$ &  34.8 & 1.87 & 4.33 & 31.0 \\
\hline
BG1  & Radial Gaussian   \space    & 0.2    & 335.74 & 60.46      &  37.5  & 2.03 & 7.00 &  36.7\\
\hline
BG1  & Radial Gaussian   \space  & 0.4    & 335.91  &  60.61        &  56.5 & 1.96 & 10.9 & 53.1  \\
\hline
Model B (two new sources) & & & & & & & & \\
\hline
G106.3+2.7 &  Point Source    \space  & None    & 336.71$^*$  & 60.90$^*$  & 19.1 & 1.65 & 1.09 & 55.6   \\
BG1  & Radial Gaussian   \space     & 0.2  & 335.71  &  60.45       &  36.1  & 2.04 & 6.92 &   \\
\hline
G106.3+2.7 &  Point Source   \space     &  None   & 336.71$^*$  & 60.90$^*$    &  14.7 & 1.62 & 0.85  &  64.4  \\
BG1  & Radial Gaussian   \space  & 0.4    & 335.70  &  60.50         & 46.9  & 1.99 & 10.4 &   \\
\hline
G106.3+2.7 &  Radial Gaussian    \space  & 0.2   & 336.65$^*$  & 60.91$^*$  & 28.0 & 1.78  & 2.97 & 62.7   \\
BG1  & Radial Gaussian   \space     & 0.2  & 335.71  &  60.45       &  33.2  & 2.05 & 6.73 &   \\
\hline
G106.3+2.7 &  Radial Gaussian   \space  & 0.2   & 336.65$^*$  & 60.91$^*$  & 18.0 & 1.72  & 2.04 & 66.7  \\
BG1  & Radial Gaussian   \space     & 0.4  & 335.74  &  60.51      &  40.4 & 2.02 & 10.1 &   \\
\end{tabular}
\end{ruledtabular}
\footnotetext[1] {Parameters fixed to the best-fit values from the $>10$~GeV analysis are denoted by an asterisk.}
\footnotetext[2] {The spectrum is assumed to follow $dN/dE_\gamma = N_0 \,(E_\gamma / 1{\,\rm GeV})^{-\alpha}$, with $N_0$ being the prefactor in units of $10^{-10}\,\rm GeV^{-1}\,cm^{-2}\,s^{-1}$ and $\alpha$ being the index. }
\end{table*}

We add a new source, G106.3+2.7, to the baseline model of the ROI, assuming that its energy spectrum follows a power-law distribution. When fitting the new ROI model to data, we free the normalization parameters of all cataloged sources within $4^\circ$ of G106.3+2.7 and the diffuse emission components. The locations of the known sources and the model parameters of sources outside the fitting circle are fixed to their best-fit values from the baseline ROI analysis. The TS for source detection is defined as twice the logarithm of the maximum likelihood values when fitting the data with and without the new source between 0.1 and 500~GeV.

Table~\ref{tab:1s_10GeV} summarizes the results that we obtained when fitting various spatial models to the full-phase data above 10~GeV. For each extended source model with a given radius, the position, flux normalization, and spectral index are left as free parameters. The extension $\sigma_{\rm ext}$ is varied from $0.15^\circ$ to $0.3^\circ$ with a $0.05^\circ$ step size. The TS peaks at  $\sigma_{\rm ext} = 0.2^\circ$ though the improvement of the TS is not significant. We also test a flat disk profile of the same radius and find it not as good as a Gaussian profile. For the fits with extended spatial templates, the statistical uncertainty of the best-fit spectral indices is $\sim 0.17$ and that of the best-fit positions is at the level of $\sim 0.05^\circ$.

Compared to the previous work \citep{Xin_2019} which uses 10 years of {\it Fermi}-LAT data above 3~GeV,  our above-$10$~GeV analysis with 12 years of data yields a less-extended, point-like emitting region with greater significance. We do not detect emission below 10~GeV or above 100~GeV.

\section{Modeling of Background Emission}\label{appendix:BG1}

\begin{figure}[t]
    \centering
    \includegraphics[width=.49\textwidth]{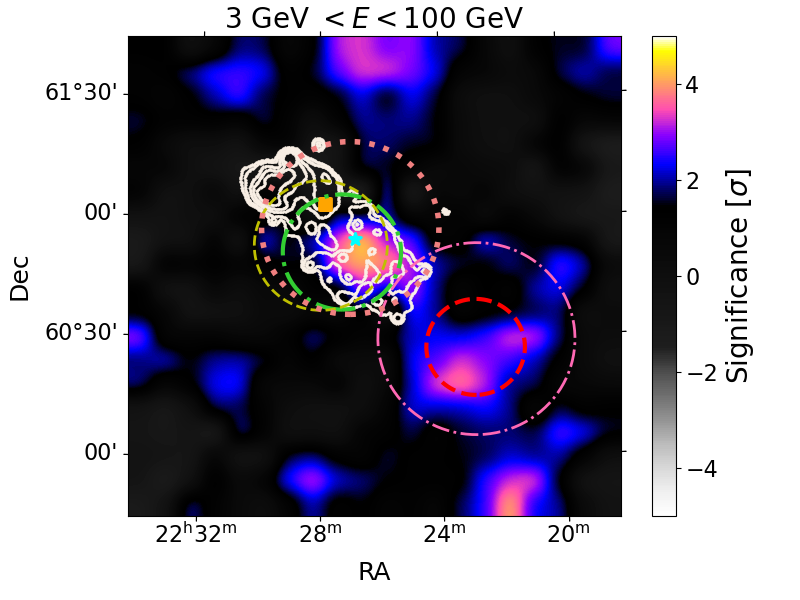}
    \caption{\label{fig:zoomedOut} Expanded-scale view of the residual map from the analysis of the phase-selected data with the weighted likelihood between 3 and 100~GeV, after fitting spectral models for known sources in the 4FGL-DR2 catalog \citep{4FGLDR2}. The best-fit position of G106.3+2.7 from the above-$10$~GeV analysis is indicated by the cyan star. The markers and contours indicating multi-wavelength observations are the same as those in Figure~\ref{fig:TSmap}. The white contours indicate the radio continuum emission of the SNR \citep{radio_2001}. The pink dash-dotted and red dashed circles indicate an extended background source BG1 with a radius of $0.4^\circ$ and $0.2^\circ$, respectively. The positions of the circles are determined by the two-source models in Table~\ref{tab:BG1} that include G106.3+2.7 as a point source. }
\end{figure}

Some nearby faint gamma-ray emission regions can be seen in the expanded-scale views of the residual excess map, as shown in Figure~\ref{fig:zoomedOut}. In particular, the excess in the southwest of G106.3+2.7 is only $0.6^\circ$ from the gamma-ray emitting site of the remnant. 
The excess could be either due to mismodeling of the Galactic diffuse emission or multiple faint sources in the same sky region, though the low statistics do not allow fitting with multiple sources. Therefore, we model the excess as one extended source, which we refer to as ``BG1".

BG1 is not detected ($\rm TS < 25$) in the full phase data above 10~GeV. When modeling it with a Gaussian profile with $0.2^\circ$ radius, we obtain $\rm TS = 11.1$. BG1 is therefore not included in the ROI model in the above-10~GeV analysis.

To examine whether BG1 is associated with G106.3+2.7 at lower energy, we compare the following two models by fitting them to the ROI: A) known 4FGL-DR2 sources and one extended source to account for the emission in the SNR and BG1 region, and B) known sources plus two independent new sources, one fixed at the best-fit position found in the above 10~GeV analysis and the other inside BG1. 
Tabel~\ref{tab:BG1} summarizes the results of the fits to the ROI between 0.1 and 500~GeV with various SNR and BG1 models. 
We find that Model~B is better than Model A with $\Delta\rm TS\sim 11-19$, depending on the assumptions of source extensions. 
Besides, the spectrum becomes softer, with the index changing from $\sim 1.7$ to $\sim 2.1$ when a test source moves from G106.3+2.7 toward BG1. 
Given that Model B is favored by data and BG1 is spatially distant from the emitting region observed by {\it Fermi}-LAT above 10~GeV and by VHE telescopes, BG1 is likely physically unrelated to G106.3+2.7.

We also test a flat disk template for BG1 and find that the disk profile provides similar fits as the Gaussian profile.  The statistical uncertainty of the best-fit spectral indices is $\sim 0.15-0.2$ and that of the best-fit positions of BG1 is at the level of $\sim 0.07^\circ$. Due to the low statistics, it is impossible to confine the extension of BG1. We therefore estimate its radius based on the extension of the cluster of emission in the residual maps above 10~GeV and above 0.1~GeV. The estimated values, $0.2^\circ$ and $0.4^\circ$, respectively, are used in the models in Table~\ref{tab:BG1}.   

\begin{figure} 
    \centering
    \includegraphics[width=.49\textwidth]{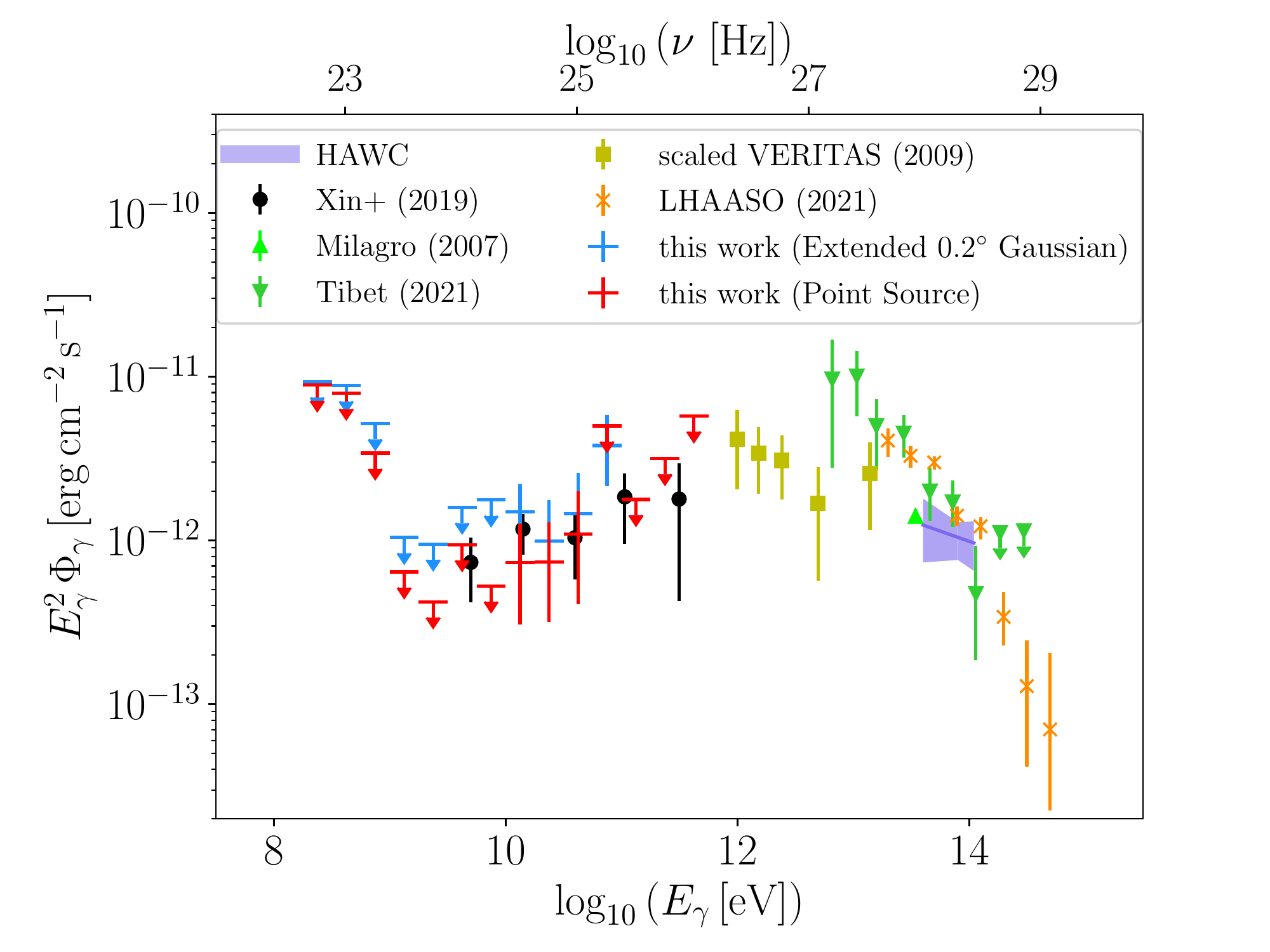}
    \caption{\label{fig:gammaSED} Spectral energy distribution (SED) of high-energy and VHE gamma-ray emission from G106.3+2.7. 
    The SEDs are obtained from the phase-gated 12-year {\it Fermi}-LAT data. 95\% upper limits are shown when $\rm TS < 4$, otherwise $1\,\sigma$ error bars are shown. The blue data points correspond to fits with the source modeled as having a Gaussian radial profile with a 68\% containment radius of $0.2^\circ$. The red data points corresponds to a conservative model where the SNR is assumed to be point-like and the nearby emitting region BG1 is assumed to be an extended source with $0.4^\circ$ radius. The last three data points in blue and red overlap. For comparison, the SED obtained by previous work \citep{Xin_2019} is shown as the black circle markers.
    The VHE flux points or band are from observations of HAWC \citep{HAWC_2020} (purple band), VERITAS \citep{Acciari_2009} (yellow square markers), Tibet AS$\gamma$ \citep{Tibet_2021} (green triangle markers), and LHAASO \citep{LHAASO_2021} (orange cross markers). The VERITAS flux points are scaled up by a factor of 1.62 from the original values \citep{Acciari_2009}.  
    }
\end{figure}

Because they are close (compared to the \textit{Fermi}-LAT PSF), the inclusion of BG1 impacts the modeling of the SNR. Among all models in Table~\ref{tab:BG1}, the case with G106.3+2.7 as a point source and BG1 as an extended source of $0.4^\circ$ radius yields the lowest flux for the SNR. Figure~\ref{fig:gammaSED} shows the SED of the SNR derived using this conservative model. The differential flux is 20--50\% lower compared to the benchmark model where the SNR is modeled as an extended source with $0.2^\circ$ radius.

\section{Multi-wavelength Observations of the Tail Region}
For observations in the radio band, we use the integrated flux of the tail region at 408~MHz and 1420~MHz by the Dominion Radio Astrophysical Observatory (DRAO) \citep{Pineault_2000}. We also include the flux density measurements of the entire SNR from the Sino-German polarization survey at 6 cm and the Effelsberg survey at 11 and 21~cm \citep{2011A&A...529A.159G}. Since the tail region was not reported separately from the head region at these higher frequencies, we estimate the flux of the tail by scaling the total flux by the tail-total flux ratio at 408~MHz. The ratio varies by 1\% from 408~MHz to 1420~MHz\citep{Pineault_2000}, so the uncertainty caused by the scaling is expected to be much smaller than the statistical uncertainty of the surveys which is $\sim 10-15\%$. 

In the X-ray band, we use the surface brightness measurement of XMM-Newton of the tail and calculate the flux using a total solid angle of 600~arcmin$^2$ \citep{2020arXiv201211531G}.  The SNR is also observed by the Suzaku satellite \citep{Suzaku_2021}. As their work \citep{Suzaku_2021} does not provide the flux of the tail, we add the surface brightness of the non-thermal model for their Middle and West fields, which is comparable to the tail of the SNR, and convert the surface brightness to an integrated flux using the Suzaku field of view of $17.8^2\,\rm arcmin^2$. The estimated Suzaku flux, indicated by the pink butterfly marker in Figure~\ref{fig:sed}, is consistent with the XMM-Newton flux. To avoid scaling errors, we do not use this estimated spectrum for broadband SED studies. % to avoid post-analysis  
%uncertainties, we approximate the intensity error of the West field using the slightly larger error of the Middle field as a conservative estimate. 

Our broadband SED fits uses the VERITAS data points below 1~TeV \citep{Acciari_2009} and the LHAASO data points above 100~TeV \citep{LHAASO_2021}. Multiple VHE gamma-ray experiments measured the spectrum between 1 and 100~TeV. We take the mean of the joint fit spectrum from VERITAS and HAWC \citep{HAWC_2020} and the spectrum from Tibet AS$\gamma$ \citep{Tibet_2021}. In each energy bin, we compare the errors in the two measurements and take the larger value as an estimate of the $1\,\sigma$ uncertainty. For our analysis we have scaled up the VERITAS flux points by a factor of 1.62 from the original values \citep{Acciari_2009} to account for the gamma-ray signals outside the extraction region, as also done in e.g. Ref.~\citep{Tibet_2021}.

For the MCMC fits, the probability function of the SED model is computed by multiplying the probabilities $p_i$ of the model in each energy bin $i$. For a bin with a detection,   $\ln\,p_i$ is defined as the squared deviation from the observed flux. For a bin with an upper limit, $p_i = 1$ if the model is below the limit and $p_i = 1 - {\rm C.L.}$ otherwise, where C.L. is the confidence level of the upper limit. 

The MCMC fits are performed with the {\it naima} software \citep{naima} using the physics models described below.  

\section{Model Selection}

The model selection is based on the likelihood-ratio test and the Bayesian information criterion (BIC). 

When models are nested, that is, when the more complex model can be transformed into the simpler model by constraining some of its parameters, a likelihood-ratio test is performed. The test statistic, defined as $-2$ times the log likelihood ratio, is assumed to follow the $\chi^2$ distribution based on Wilks' theorem \citep{10.1214/aoms/1177732360}. The probability of rejecting the null hypothesis is converted to a significance level using the normal distribution. 

In more general cases, we calculate the BIC value. BIC is defined as $\mathrm {BIC} =k\ln(n)-2\ln(\hat{\cal {L}})$, where $k$ is number of parameters used by the model, $n$ is the number of data points in use, and $\hat{\cal{L}}$ is the maximized value of the likelihood function. Models with lower BIC values are generally preferred and a difference in BIC of greater than 10 means that the evidence favoring the better model is very strong \cite{BICbook}.

\section{Leptonic Models}\label{appendix:lep}

Depending on their energy, the cooling time of relativistic electrons may be shorter or longer than the age of the SNR. We model the evolution of the electron spectrum with a transport equation 
\begin{equation}\label{eqn:transport}
    \frac{\partial N_e}{\partial t} + \frac{\partial}{\partial \gamma_e}\left[\dot{\gamma_e}\,N_e(\gamma_e, t)\right] = Q_e(\gamma_e, t), 
\end{equation}
where $\dot{\gamma}_e = -{4}/{3}\,\gamma_e^2\, c\, \sigma_T\,(u_B + u_{\rm \gamma}\,F_{\rm KN}(\gamma_e)) / (m_e\,c^2)$ is the energy loss rate due to inverse-Compton and synchrotron emission, $\sigma_T$ is the Thomson cross section, $F_{\rm KN}$ is the $\gamma_e$-dependent cross section suppression factor due to the Klein-Nishina effect \cite{2005MNRAS.363..954M}, $N_e = dN_e/dE_e$ and $Q_e=dN_{e0}/dE_edt$ are the spectrum and injection rate of electrons, and $u_B = B^2 / (8\pi)$ and $u_{\gamma}$ are the energy density of magnetic field and the radiation field, respectively. 

We assume that electrons are injected constantly over time and solve equation~\ref{eqn:transport} for every set of free parameters $(N_{e0}, \alpha_e, E_{e,\rm max}, B)$ that is sampled by the MCMC. For two-component leptonic models we solve two transport equations that share the magnetic and radiation background of the SNR but have different injection spectra. The steady-state electron spectrum is then used to compute the inverse-Compton and synchrotron fluxes that are later compared with the multi-wavelength data. The electron injection rate may evolve over time and be impacted by the dynamical evolution of the nebula, though the corresponding effect on the $\gamma$-ray spectral shape is not strong if the nebula is in the free expansion phase (see e.g. \cite{2022MNRAS.tmp...45F}).

\begin{figure} 
    \centering
     \includegraphics[width=.49\textwidth]{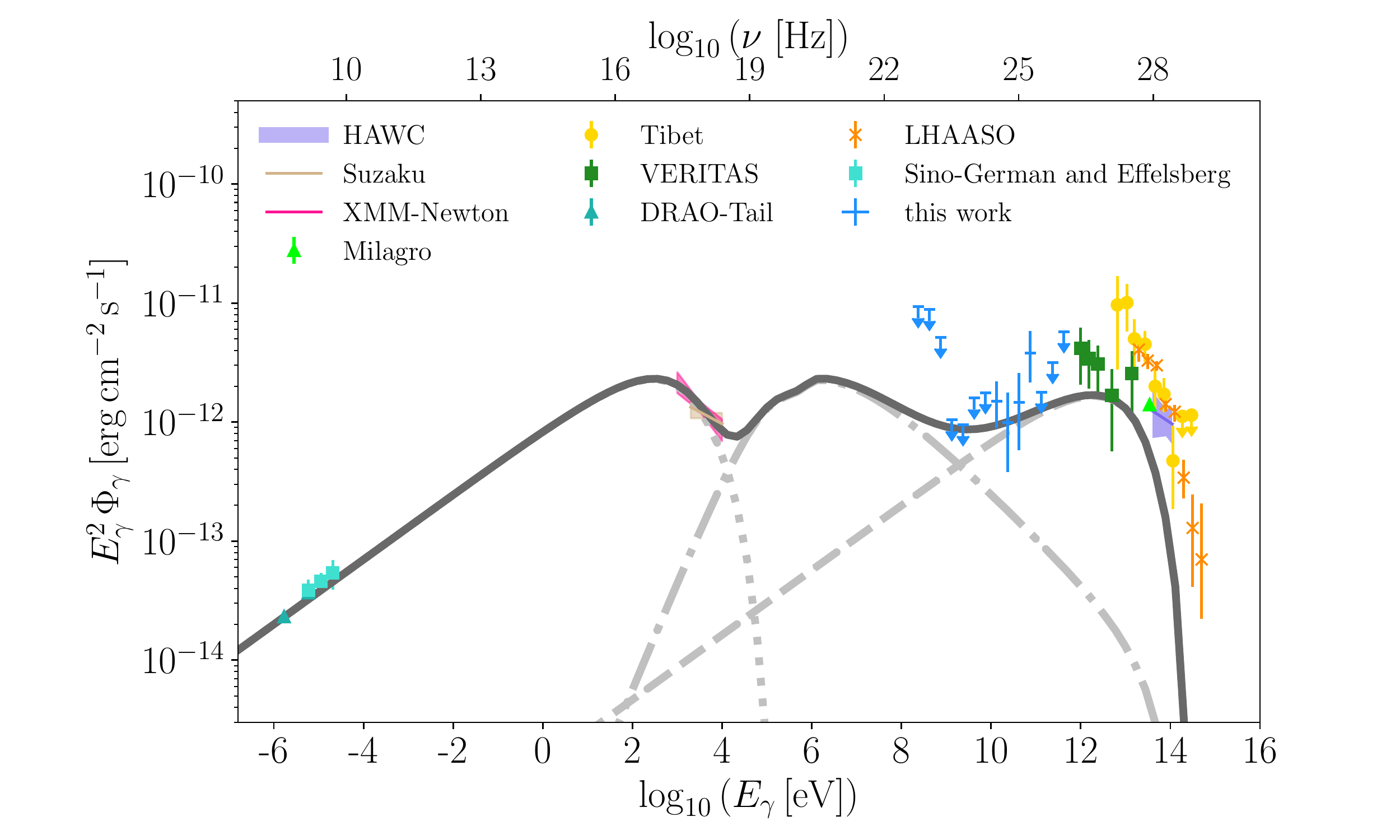}
     \includegraphics[width=.49\textwidth]{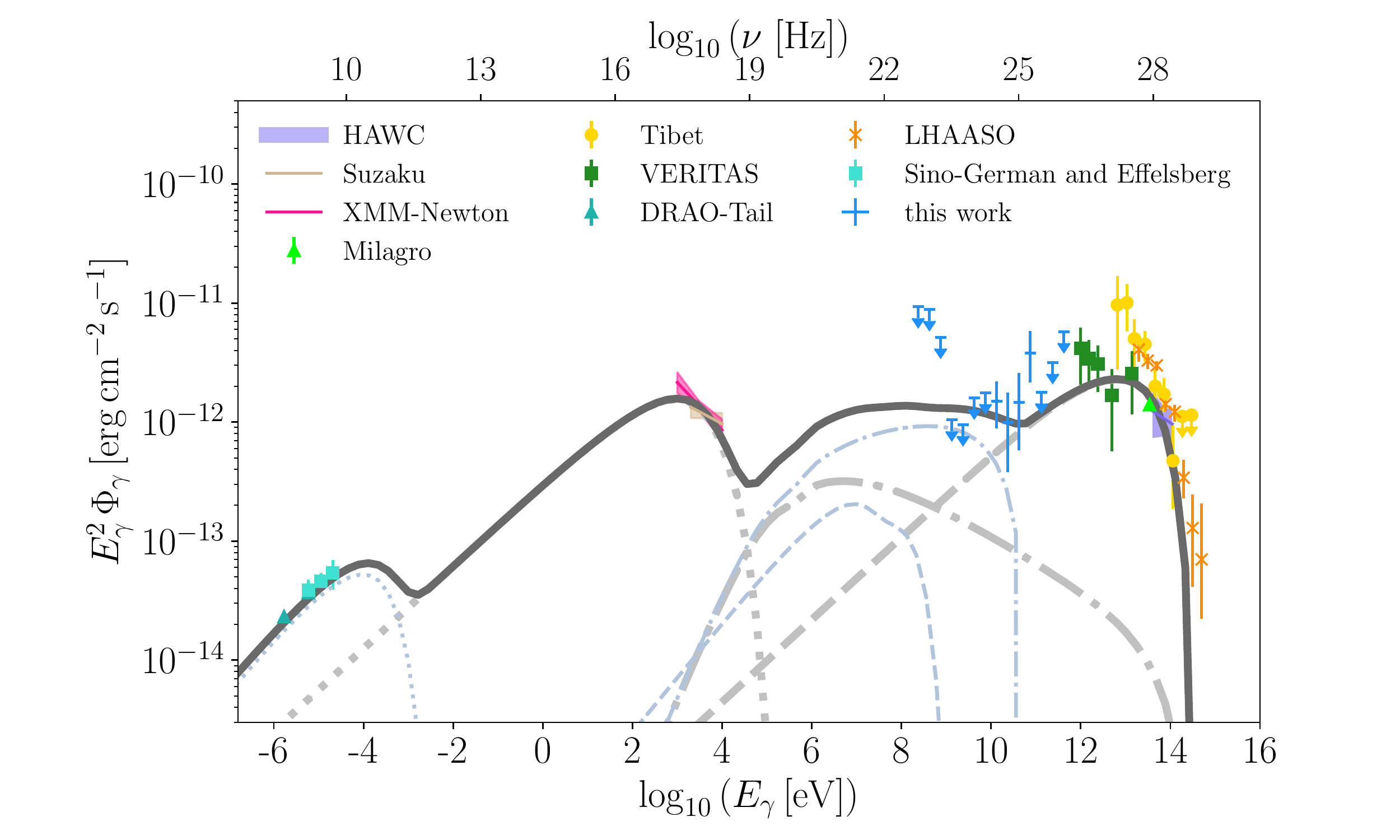}
    \caption{\label{fig:sed_supp} Leptonic SED models of G106.3+2.7. The multi-wavelength data is the same as in Figure~\ref{fig:sed}. The VERITAS flux points are again scaled up by a factor of 1.62 from the original values \citep{Acciari_2009, Tibet_2021}. The top plot shows the one-component leptonic model where a single electron population produces emission in all wavebands. The best-fit parameters include $W_e=3.4\times10^{48}$~erg, $\alpha_e = 2.5$, $E_{e\,\rm max}= 2.9\times 10^{14}$~eV, and $B=3.9\,\mu$G. The bottom panel shows the two-component model where a low-energy electron population (light blue) produces the radio and MeV-to-GeV gamma-ray emission and a high-energy electron population (grey) produces the X-ray and VHE gamma-ray emission. The best-fit parameters are $W_{e1}=4.4\times10^{47}$~erg, $\alpha_{e1} = 2.3$, $E_{e1\,\rm max}= 3.8\times 10^{14}$~eV, $W_{e2}=7.8\times10^{47}$~erg, $\alpha_{e2} = 2.1$, $E_{e2\,\rm max}= 5.4\times10^{10}$~eV, and $B=2.2\,\mu$G. The Synchrotron, Bremsstrahlung, inverse Compton components, and their sum are indicated by dotted, dash-dotted, dashed, and solid curves, respectively.   
    } 
\end{figure}

Figure~\ref{fig:sed_supp} presents the best-fit one-component and two-component leptonic models based on the {\it Fermi}-LAT flux measurement using an extended source model. With three more free parameters, the two-component leptonic model provides a better fit to the data than the one-component leptonic model with ${\rm TS} = 11.6$ and $\rm (BIC)_{1e} - (BIC)_{2e} = 0.5$. The improvement is insufficient to justify the three extra degrees of freedom. The two-component leptonic model is disfavored relative to the lepto-hadronic model with $\rm (BIC)_{2e} - (BIC)_{had} = 20.1$ as explained in the main text.

When using the {\it Fermi}-LAT SED obtained from the conservative point-source model, the leptonic models are further disfavored. We find a difference in BIC values $\rm (BIC)_{e} - (BIC)_{had} = 30.8$ for the one-component leptonic model and the lepto-hadronic model. Since these two models are nested, we can also compare their TS. The TS of the lepto-hadronic model is higher by $\Delta \rm TS = 43.7$, which corresponds to $5.9\,\sigma$. The lepto-hadronic model is also significantly better than the two-component leptonic model with $\rm (BIC)_{2e} - (BIC)_{had} = 40.7$.
Since the difference in BIC values is $\gg 10$, the strength of the evidence against the leptonic models is very strong.

In equation~\ref{eqn:transport} we have ignored the diffusion of electrons as the diffusion time is likely longer than the cooling time of VHE electrons or the source age, and also because doing so saves computing time. Due to the presence of turbulent magnetic field at the acceleration site, the diffusion coefficient is usually much smaller than the average diffusion coefficient in the ISM, $D_{\rm ISM}\approx 10^{28}\,(E/3\,\rm GeV)^\delta$, with $\delta\sim 0.33$ for the  Kolmogorov turbulence \cite{1990acr..book.....B}.  Had this not been the case, and if the gamma rays were produced by electrons, the size of the gamma-ray emitting region, $R\sim (2D_{\rm ISM}t_{\rm age})^{1/2}= 67\,{\rm pc} (E/ 1\,\rm TeV)^{\delta/2}$, would have been much more extended than what has been observed, $R_{\rm SN}\lesssim 10\,\rm pc$.   

We have assumed that both hadronic cosmic rays and the ISM gas are protons, though the composition could be heavier. For example, the gas number density could be $1/\mu\sim 1.6$ times higher, where $\mu\approx 0.62$ is the mean molecular weight of fully ionized gas of solar composition \citep{zombeck_2006}. The cross section for the production of pions would also increase to $\sim (A_{\rm CR}\,A_T)^{2/3}$ times higher, with $A_{\rm CR}$ and $A_T$ being the mass numbers of the projectile cosmic ray and the target, respectively  (e.g.,\citep{2007NIMPB.254..187N}). The effects of a heavier composition include a decrease of $W_p$ in the hybrid model and an increase of the Bremsstrahlung emission by electrons, which makes leptonic models even less favored. The maximum energy of hadronic cosmic rays would be $\sim A_{\rm CR}\,E_{p,\rm max}$ to explain the observed maximum $\gamma$-ray energy.

\section{Lepto-hadronic Hybrid Models}

As the cooling time of protons is much longer than the source age, $t_{\rm pp}\approx 1/(n_{\rm gas}\,\sigma_{\rm pp}c) = 20\,\rm Myr\,(n_{\rm gas}/1\,\rm cm^{-3})^{-1} \gg t_{\rm age}$, where  $\sigma_{\rm pp}\approx 50\,\rm mb$ is the proton-proton inelastic cross section around 100~TeV \cite{Olive_2016}, the injection history of protons barely impacts the gamma-ray spectrum. The injection spectrum $dN_p/dE_p$ is therefore used to compute the pion production using the cross sections from Ref.~\cite{2014PhRvD..90l3014K}.

Secondary electrons are produced by the decay of charged pions, though their relative contribution to the gamma-ray emission is negligible. The total energy of secondary electrons produced over the source lifetime can be estimated as $W_e^{\rm sec} \approx W_p n_{\rm gas} \sigma_{\rm pp} c t_{\rm age} = 1.6\times 10^{45}\,{\rm erg}\,(W_p/4\times 10^{48}\,\rm erg)\,(n_{\rm gas}/1\,\rm cm^{-3})\,(t_{\rm age}/10\,\rm kyr)$, where $W_p$ is the energy carried by primary protons. $W_e^{\rm sec}$ is much smaller than the best-fit electron energy $W_e = 5.3\times 10^{47}\,\rm erg$, suggesting that the synchrotron emission is dominantly produced by primary electrons.      

The best-fit proton spectral index is harder than that traditionally associated with diffusive shock acceleration. This could be caused by either an increase in the shock compression ratio due to the presence of relativistic particles \citep{1984A&A...132...97A, 2011JCAP...05..026C} when the shock acceleration efficiency is high, or by the very highest energy particles escaping ahead of the shock front \citep{2005MNRAS.361..907B}. A hard gamma-ray spectrum may also be explained by a scenario, initially proposed for SNR RX~J1713.7$-$3946  \cite{2010ApJ...708..965Z, 2012ApJ...744...71I, 2014MNRAS.445L..70G, 2019MNRAS.487.3199C}, where cosmic rays accelerated by the shocks penetrate a dense gas clump. Since the diffusion coefficient is an increasing function of particle energy, higher-energy particles penetrate more effectively, leading to a harder spectrum than at the shock. 

%{\st {Interestingly, with such a hard spectrum, a minority of the SNR population would be sufficient to explain the PeV cosmic-ray intensity.}} 

\end{document}